\documentclass[journal]{IEEEtran}
\usepackage{cite}
\usepackage{lineno,hyperref}

\usepackage{graphicx}
\usepackage{bm}
\usepackage{subfigure}

\ifCLASSINFOpdf
  \usepackage{graphicx}
  \graphicspath{{../pdf/}{../jpeg/}}
  \DeclareGraphicsExtensions{.pdf,.jpeg,.png}
\else
\fi
\begin{document}
%
\title{Robust adaptive steganography based on dither modulation and modification with re-compression}
%
%
%

\author{Zhaoxia~Yin,~\IEEEmembership{Member,~IEEE,}
        Longfei~Ke~\IEEEmembership{}
\thanks{Zhaoxia~Yin and Longfei~Ke are with Anhui Province Key Laboratory of Multimodal Cognitive Computation, School of Computer Science and Technology, Anhui University, 230601, CHINA e-mail:yinzhaoxia@ahu.edu.cn.}
\thanks{Manuscript received April 19, 2005; revised August 26, 2015.}}

%
%

\markboth{Journal of \LaTeX\ Class Files,~Vol.~14, No.~8, August~2015}%
{Shell \MakeLowercase{\textit{et al.}}: Bare Demo of IEEEtran.cls for IEEE Journals}
%



\maketitle

\begin{abstract}
Traditional adaptive steganography is a technique used for covert communication with high security, but it is invalid in the case of stego images are sent to legal receivers over networks which is lossy, such as JPEG compression of channels. To deal with such problem, robust adaptive steganography is proposed to enable the receiver to extract secret messages from the damaged stego images. Previous works utilize reverse engineering and compression-resistant domain constructing to implement robust adaptive steganography. In this paper, we adopt modification with re-compression scheme to improve the robustness of stego sequences in stego images. To balance security and robustness, we move the embedding domain to the low frequency region of DCT (Discrete Cosine Transform) coefficients to improve the security of robust adaptive steganography. In addition, we add additional check codes to further reduce the average extraction error rate based on the framework of E-DMAS (Enhancing Dither Modulation based robust Adaptive Steganography). Compared with GMAS (Generalized dither Modulation based robust Adaptive Steganography) and E-DMAS, experiment results show that our scheme can achieve strong robustness and improve the security of robust adaptive steganography greatly when the channel quality factor is known.
\end{abstract}


%
\IEEEpeerreviewmaketitle

\section{Introduction}
%
%
%
%

\IEEEPARstart{T}{o} achieve copyright protection, covert communication and other functions, data hiding is a technology which embeds additional data into digital multimedia \cite{zhang2017Data}, such as audio \cite{wu2020audio}, video \cite{yao2021motion}, 2D vector graphics \cite{2019Reversible}, 3D mesh models \cite{jiang2017reversible}, images \cite{fan180reversible} and so on.  In data hiding, there is a development trend of multi-field combine, \cite{2020Separable} proposed a data hiding scheme with robustness, reversibility and encryption. Among them, focusing on covert communication and detection, steganography and steganalysis \cite{wang2004cyber} have made considerable progress. Nowadays, the most popular image steganography method is adaptive steganography scheme which defines a distortion function to calculate modification costs of all elements, then we can embed secret messages into cover elements with Syndrome-Trellis Codes (STCs) \cite{filler2011minimizing}, such as J-UNIWARD (JPEG UNIversal WAvelet Relative Distortion) \cite{holub2014universal}, UERD (Uniform Embedding Revisited Distortion) \cite{guo2015using} in JPEG domain. These schemes \cite{holub2014universal,guo2015using} are only suitable for laboratory environment which assumes the receiver can receive the stego images losslessly. 
\par Nowadays, a large amount of data is sent over networks to achieve the purpose of interaction, but some factors such as data loss, additional noise and attacks in the networks cause the received data unreliable. Legal receivers need to process the received data in order to restore the original state of received data as much as possible. For example, \cite{8247236} presents a data fusion scheme to reduce false data injection attack's effects on a networked radar system. \cite{7563348} presents a distributed weighted average consensus scheme which is robust to data falsification attacks in distributed networks. \cite{8859263} proposes a fast max-based consensus scheme which robust to additive noise in wireless channels.
\par With the development of social networks, lots of people share their life photos over social networks. Combining steganography with this common behavior to embed secret messages into the shared photos, which hides the behavior of covert communication. However, current social networks, such as Facebook, execute lossy process (JPEG compression and scaling) on the shared images because of limited bandwidth and storage. The shared images embedded with secret messages will be destroyed irreversibly when it be sent to legal receivers over networks, and the legal receivers can not recover the secret secret messages from the received images correctly. In order to apply steganography to real life, it is necessary to improve the robustness of adaptive steganography scheme. Therefore, robust adaptive steganography algorithms applied to lossy channel have to possess strong detection resistant capability (Security) and robust to the lossy operation of the channel (Robustness). Security refers to the possibility of stego images not be discovered by the steganalysis algorithm, and robustness ensures that the legal receiver can correctly extract secret messages. This article focuses on the JPEG-compression of channel when images are sent over networks. 
\par Currently, several studies tried to achieve robust adaptive steganography. In GMAS \cite{yu2020robust}, the robust adaptive steganography is divided into three categories according to the application: 1) "Upward Robust", 2) "Downward Robust", 3) "Matching Robust".
\par "Upward Robust" is the schemes work in the quality factor of channel JPEG compression not smaller than the quality factor of cover images. In this mode, Zhang et al. proposed a structure of "Compression-resistant Domain Constructing + RS + STCs Codes" \cite{zhang2016framework}. Based on the framework, Zhang et al. proposed several works such as DCRAS (DCT Coefficients Relationship based Adaptive Steganography ) \cite{zhang2015jpeg}, FRAS (Feature Region based Adaptive Steganography) \cite{zhang2017joint}, DMAS (Dither Modulation based robust Adaptive Steganography) \cite{zhang2018dither}. DCRAS \cite{zhang2015jpeg} utilizes the robustness of the relationship between DCT coefficients to embed messages. FRAS \cite{zhang2017joint} obtains robust embedding regions based on feature region extraction and selection algorithm. DMAS \cite{zhang2018dither} modifies middle frequency DCT coefficients based on dither modulation algorithm to embed secret messages. Although these schemes \cite{zhang2015jpeg,zhang2017joint,zhang2018dither} can achieve high extraction accuracy, but weak security of the schemes is a serious disadvantage. Yu et al. proposed GMAS \cite{yu2020robust} based on DMAS \cite{zhang2018dither}. They replace symmetric distortion with asymmetric distortion, combine with ternary STCs and expand the embedding domain to the lower frequency regions. GMAS \cite{yu2020robust} can achieve strong robustness, especially, when the channel quality factor $Q_{c}$ is known, they can achieve better performance by selecting cover images with quality factor $Q_{c}$. However, GMAS \cite{yu2020robust} adopts the method of encoding secret messages with RS (Reed-Solomon) codes \cite{macwilliams1977theory}, which means with the increase of payload, the security will decline rapidly because of the large number of check codes. These schemes \cite{yu2020robust,zhang2015jpeg,zhang2017joint,zhang2018dither} embed secret messages with STCs \cite{filler2011minimizing}, but errors in stego sequences extracted from damaged stego images will appear error diffusion phenomenon in the STCs decoding \cite{liu2015damage,zhang2018fault}. Based on these researches, Zhang et al. proposed a new framework "Compression-resistant Domain Constructing + STC + CRC Codes" \cite{zhang2020enhancing}, which adopts CRC (Cyclic Redundancy Code) codes \cite{1311885} to encode the stego sequences for fewer check codes. Zhang et al. modify DMAS \cite{zhang2018dither} with the framework and get a scheme called E-DMAS. The scheme can achieve higher security than DMAS at high payloads. E-DMAS \cite{zhang2020enhancing} can solve the problem of rapid decreasing in security caused by embedding a large number of check codes. But the same problem as DMAS \cite{zhang2018dither}, secret messages are embedded into middle frequency DCT coefficients, security is still an issue, although better than DMAS \cite{zhang2018dither}. 
\par "Downward Robust" is the schemes work in the quality factor of channel JPEG compression smaller than the quality factor of cover images. Tao et al. proposed a robust image steganography scheme for the situation in \cite{8533349}. At first, they re-compress an original image with channel quality factor and get a cover image, then they embed a secret message into the cover image with J-UNIWARD \cite{holub2014universal} or UERD \cite{guo2015using} to get a stego image. Then, they modify DCT coefficients of the original image according to the stego image and get an intermediate image, so that the intermediate image is compressed with channel quality factor to obtain the stego image. The scheme can extract the secret message from the stego image completely and achieve high security after channel JPEG compression. But the DCT coefficient residuals of the intermediate image and the corresponding original image are too large to guarantee security. In \cite{8901147}, Zhu et al. utilize the robustness of DCT coefficient sign to implement robust adaptive steganography.
\par "Matching Robust" is the schemes which re-compress cover images multiple times to reduce impact of social network. In \cite{8566016}, Zhao et al. proposed a robust adaptive steganography scheme based on transport channel matching. They re-compress a cover image multiple times with the channel quality factor before embedding a secret message. They encode the secret message with BCH code \cite{1053825} to improve the extraction accuracy of the secret message. The scheme can achieve strong robustness and high security. Besides, the capacity of the scheme is large. But it is suspicious behavior and time-consuming.
\par Based on the framework proposed in \cite{zhang2020enhancing}, we mitigate the security shortcomings of GMAS \cite{yu2020robust} and E-DMAS \cite{zhang2020enhancing}. We adopt modification with re-compression scheme to improve robustness, which provide possibility to move the embedding domain to low frequency regions of DCT coefficients and pursue trade-off between robustness and security. The experimental results show that the proposed scheme can achieve same (or stronger) robustness and higher security (especially under higher payloads) than GMAS \cite{yu2020robust} and E-DMAS \cite{zhang2020enhancing} when the channel JPEG compression quality factor is utilized.  
\par The main contributions of this paper are listed as follow:\\
(1) Combining with the known channel compression quality factor, we propose a modification with re-compression scheme based on the framework proposed in \cite{zhang2020enhancing} to reduce errors in the extracted stego sequences. \\
(2) Based on the framework proposed in \cite{zhang2020enhancing}, we adopt additional check codes to improve the robustness of steganography. Besides, we balance robustness and security by moving embedding domain to lower frequency regions to improve security of our scheme. 
\par In the next section of this paper, we will introduce related works in Section II. The proposed method will be introduced in Section III. The experiment results and discussion are shown in Section IV. Conclusion is listed in Section V.


 

\section{Related works}
\label{sec::Proposed method}
\par In this part, the basic idea of dither modulation algorithm, GMAS algorithm \cite{yu2020robust} and E-DMAS algorithm \cite{zhang2020enhancing} are introduced. For convenience, we represent embedding domains with corresponding symbols in this part.

\subsection{Notations}
\label{subsec::room reservation}
\par In this paper, all matrices and vectors are represented in bold. $\bm{X}=(x_{ij})^{n_{1}\times n_{2} },\bm{Y}=(y_{ij})^{n_{1}\times n_{2} }$ represent a cover image and a stego image with size $n_{1}\times n_{2}$ respectively. All cover and stego images in this paper are JPEG images. The symbol $\bm{J^{-1}(\bm{X})} $ represent the image that $\bm{X}$ is decompressed to spatial domain.

\subsection{Representation of embedding domain}
\begin{figure}[!ht]
	\centering
	\includegraphics[width=0.25\textwidth]{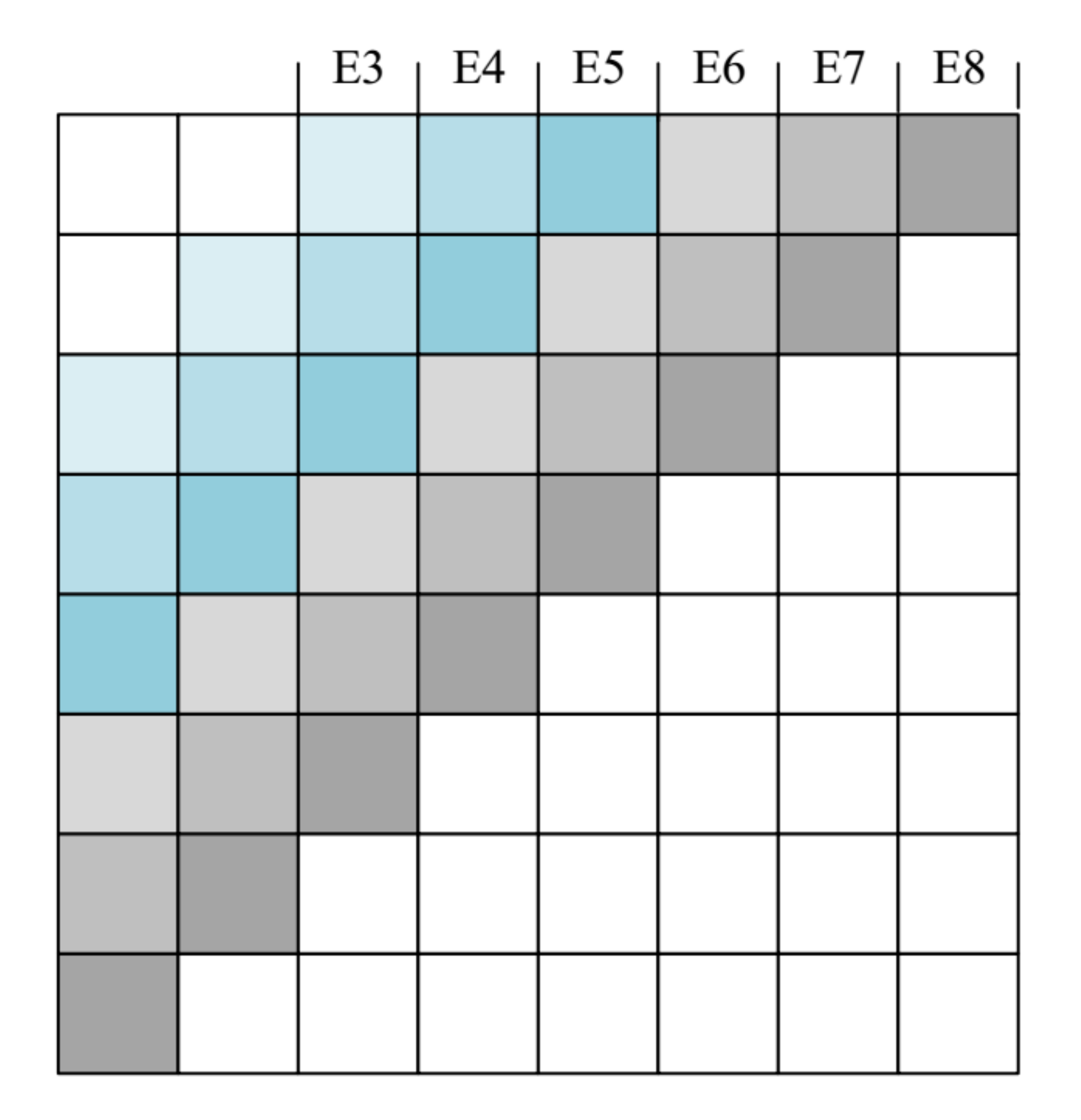}
	\caption{Representation and symbol of embedded domain in a $8\times 8$ DCT block}
	\label{fig1}
\end{figure}

\par We represent different embedding domains in a $8\times 8$ DCT coefficient block with different symbols, for example, E8 represents the 8 DCT coefficients in counter-diagonal with the same color as shown in Fig.1. And related embedding domain will be represented as E\_number1+number2. For example, E\_78 represents combination of E7 and E8.
\subsection{Dither modulation algorithm}
\begin{figure}[!ht]
	\centering
	\includegraphics[width=0.4\textwidth]{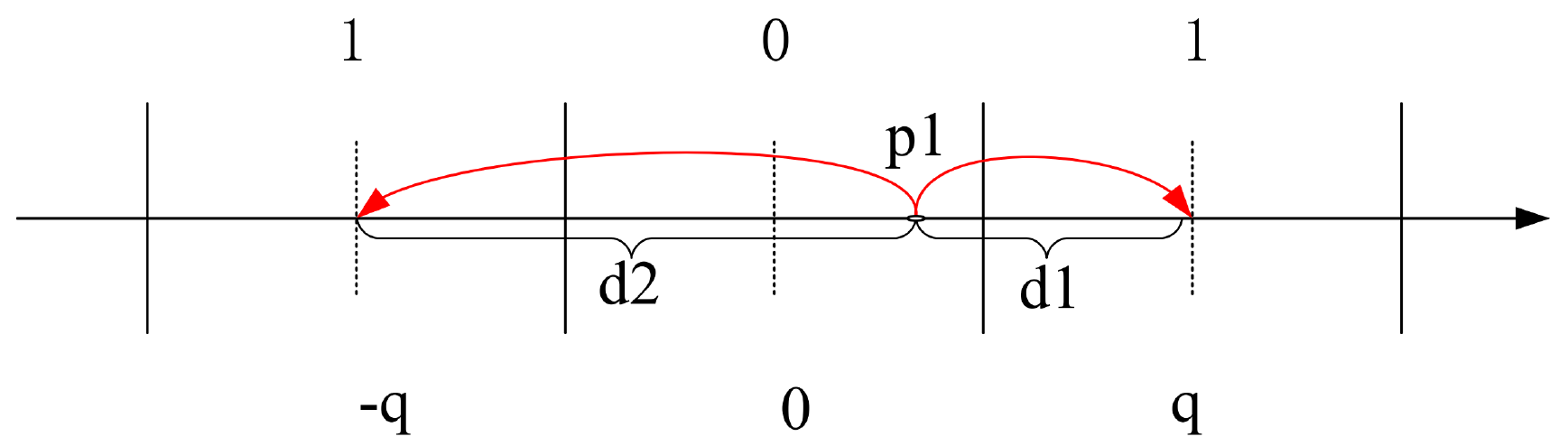}
	\caption{Embedding scheme of dither modulation algorithm}
	\label{fig2}
\end{figure}
\par Dither modulation is an implementation of Quantization Index Modulation (QIM)\cite{866336,noda2006high} watermarking scheme. Just as shown in Fig.2, according to the quantization step $q$, the coordinate axis is divided into multiple intervals. In the dither modulation algorithm, if a de-quantized DCT coefficient value on the coordinate axis belong to an odd interval, which represents a cover element "1", otherwise represents "0". Message "1" is embedded into a de-quantized DCT coefficient which belong to an even interval, the de-quantized DCT coefficient will be modified to middle coordinate of the nearest odd interval, just like the point $p1$ and modification distance is $d1$ in Fig.2. It is similar when it embeds "0" into a de-quantized DCT coefficient belong to odd interval. 
\par In the generalized dither modulation algorithm proposed in GMAS \cite{yu2020robust}, cover elements are quantized DCT coefficient values. When it embeds message "1" into a de-quantized DCT coefficient which belong to an even interval, the de-quantized DCT coefficient can be modified to middle coordinate of the two nearest odd intervals just like the point $p1$ and modification distances are $d1$ and $d2$ in Fig.2. It is similar when it embeds "0" into a de-quantized DCT coefficient belong to an odd interval.

\subsection{Review of GMAS}
\label{subsec::encrypted image}
\par Yu et al. propose GMAS \cite{yu2020robust} with a improved embedding method and embedding region based on DMAS \cite{zhang2018dither}. They achieve a wonderful trade-off between robustness and security base on the framework proposed in \cite{zhang2016framework}. The basic idea of GMAS will be introduced briefly as follow.\\
1. For a given cover image $\bm{X}$, calculate the de-quantized DCT coefficients of the cover image.\\
2. Calculate embedding distortion $\bm{\rho}$ of all cover elements with distortion function (e.g. J-UNIWARD) according to Eq.1. $W_{\mu\nu}^{(k)}$ is $uv$th wavelet coefficient in the $k$th subband of the first decomposition level, and $\sigma=2^{-6}$.\\
\begin{equation}
\rho_{ij}=\sum\limits_{k=1}^{3}\sum\limits_{\mu=1}^{n_{1}}\sum\limits_{\nu=1}^{n_{2}}\frac{|W_{\mu\nu}^{(k)}(J^{-1}(X))-W_{\mu\nu}^{(k)}(J^{-1}(Y_{x_{ij}}))|}{|W_{\mu\nu}^{(k)}(J^{-1}(X))|+\sigma}
\label{eq1}
\end{equation}
3. Calculate asymmetric distortion $\bm{\rho}^{+}$, $\bm{\rho}^{-}$ according to Eq.2 and Eq.3. $\overline{x_{ij}}$ represents the de-quantized DCT coefficients of the processed image, $\alpha \in [0,1]$. \\
\begin{equation}
\rho_{ij}^{+}=
\left\{
\begin{array}{lr}
\alpha\cdot \rho_{ij}  , & x_{ij}<\frac{\overline{x_{ij}}}{q_{ij}}\\
\rho_{ij} , & x_{ij}\geq  \frac{\overline{x_{ij}}}{q_{ij}}\\
\end{array}
\right.
\label{eq2}
\end{equation}
\begin{equation}
\rho_{ij}^{-}=
\left\{
\begin{array}{lr}
\alpha\cdot \rho_{ij}  , & x_{ij}>\frac{\overline{x_{ij}}}{q_{ij}}\\
\rho_{ij} , & x_{ij}\leq  \frac{\overline{x_{ij}}}{q_{ij}}\\
\end{array}
\right.
\label{eq3}
\end{equation}
4. Extract a cover sequence $\bm{C}$ and the modification distances $\bm{d}^{+}$, $\bm{d}^{-}$ from E\_678 of all $8\times 8$ DCT blocks with generalized dither modulation algorithm.\\
5. Calculate modifying costs $\bm{\xi}^{+}, \bm{\xi}^{-}$. $\rho_{ij}$ represents the cover image distortion of  $ij$th quantized DCT coefficient, $q_{ij}$ is quantization step. According to Eq.4 and Eq.5, $\zeta_{ij}$ represents the cover image distortion of  $ij$th de-quantized DCT coefficient, modifying costs $\bm{\xi}^{+}, \bm{\xi}^{-}$ can be obtained with Eq.4 and Eq.5 respectively.\\
\begin{equation}
\zeta_{ij}^{+}=\frac{\rho_{ij}^{+}}{q_{ij}},\xi_{ij}^{+}=\zeta_{ij}^{+}\times d_{ij}^{+}
\label{eq4}
\end{equation}
\begin{equation}
\zeta_{ij}^{-}=\frac{\rho_{ij}^{-}}{q_{ij}},\xi_{ij}^{-}=\zeta_{ij}^{-}\times d_{ij}^{-}
\label{eq5}
\end{equation}
6. Encode a secret massage $\bm{m}$ with RS codes to get a encoded massage $\bm{m'}$.\\
7. Embed the encoded message $\bm{m'}$ into the cover sequence  $\bm{C}$ with ternary STCs to get a stego sequence $\bm{S}$. And quantize the de-quantized DCT coefficients of the cover image with the stego sequence $\bm{S}$ and modification distances $\bm{d}^{+}$,  $\bm{d}^{-}$. A stego image $\bm{Y}$ can be obtained with the quantized DCT coefficients.
\par The receiver uses the same quantization table to calculate the quantized DCT coefficients of the received stego image, then, the receiver utilize STCs to decode the stego sequence extracted from the received stego image, and extract the secret message with RS decoding.
\subsection{Review of E-DMAS}
\par Zhang et al. proposed a new framework of robust adaptive steganography in \cite{zhang2020enhancing}. The framework requires cover sequences extracted from cover images are robust. Because CRC codes \cite{1311885} has limited error correction capability. Besides, the length of the check codes is related to the length of the cover sequences, with the length of the check codes increases, the probability of being detected by the steganalysis algorithms increases, therefore, the length of the cover sequences should not be too long. The basic idea of E-DMAS is introduced briefly as follow.\\
1. For a given cover image $\bm{X}$, calculate the de-quantized DCT coefficients of the cover image.\\
2. Calculate embedding distortion $\bm{\rho}$ of all cover elements with distortion functions (e.g. J-UNIWARD) according to Eq.1. \\
3. Extract a cover sequence $\bm{C}$ and corresponding modification distances $\bm{d}$ from E\_78 of all $8\times 8$ DCT blocks with dither modulation algorithm. \\
4. Calculate modification costs $\bm{\xi}$, $\rho_{ij}$ represents the cover image distortion of  $ij$th quantized DCT coefficient, $q_{ij}$ is quantization step. According to Eq.6, $\zeta_{ij}$ represents the cover image distortion of  $ij$th de-quantized DCT coefficient, modifying costs $\bm{\xi}$ can be obtained with Eq.6.\\
\begin{equation}
\zeta_{ij}=\frac{\rho_{ij}}{q_{ij}},\xi_{ij}=\zeta_{ij}\times d_{ij}
\label{eq6}
\end{equation}
5. Scramble the cover sequence $\bm{C}$ to get scrambled cover sequence $\bm{C'}$, embed a secret message $\bm{m}$ into first $l_{e}$ bits of the scrambled cover sequence $\bm{C'}$ with STCs and obtain a stego sequence $\bm{S1}$. $l_{e}$ can be calculated with Eq.7, $l_{c}$ represents the length of the cover sequence, $l_{r}$ represents the length of the message in each group CRC codes, $k$ is the highest power of generator polynomial.\\
\begin{equation}
l_{e}=l_{c}-\lceil\frac{l_{c}}{l_{r}}\rceil\cdot k
\label{eq7}
\end{equation}
6. Encode the stego sequence $\bm{S1}$ with CRC codes \cite{1311885}, and the check codes are embedded into the rest $l_{c}-l_{e} $ bits scrambled cover sequence to obtain a stego sequence $\bm{S2}$. Then, we can get a stego sequence $\bm{S}$ which is composed of $\bm{S1}$ and  $\bm{S2}$.\\
7. Inverse scramble the stego sequence $\bm{S}$ and get a sequence $\bm{S'}$. De-quantized DCT coefficients of the cover image are modified with the stego sequence $\bm{S'}$, so that the sequence extracted from the modified de-quantized DCT coefficients is consistent with the stego sequence $\bm{S'}$. Finally, a stego image $\bm{Y}$ can be obtained with the modified DCT coefficients.
\par The receiver calculates the de-quantized DCT coefficients of the received stego image, then, the stego sequence is extracted with dither modulation. Check codes are extracted from the last $l_{c}-l_{e} $ bits scrambled stego sequence with STCs decoding to correct the first $l_{e} $ bits. After that, the secret message is extracted from the corrected first $l_{e} $ bits scrambled stego sequence.
\section{Proposed method}
\subsection{Motivation}
\begin{figure}[!ht]
	\centering
	\includegraphics[width=0.4\textwidth]{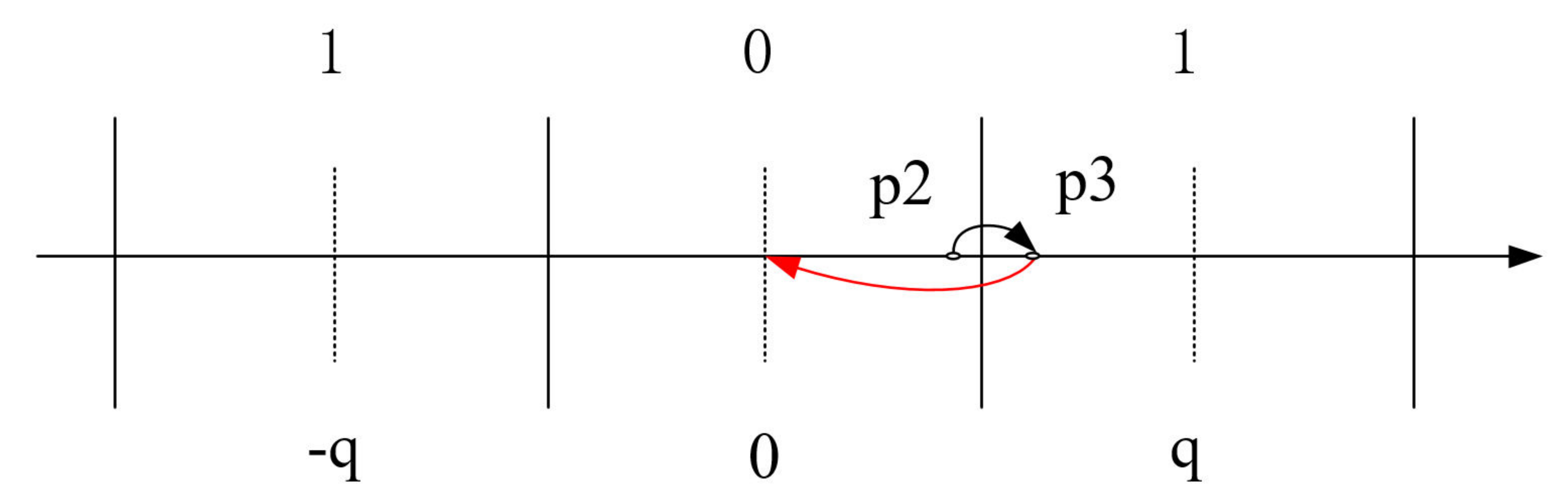}
	\caption{Modification scheme of unstable DCT coefficients}
	\label{fig3}
\end{figure}
\begin{figure*}[!ht]
	\centering
	\includegraphics[width=0.9\textwidth]{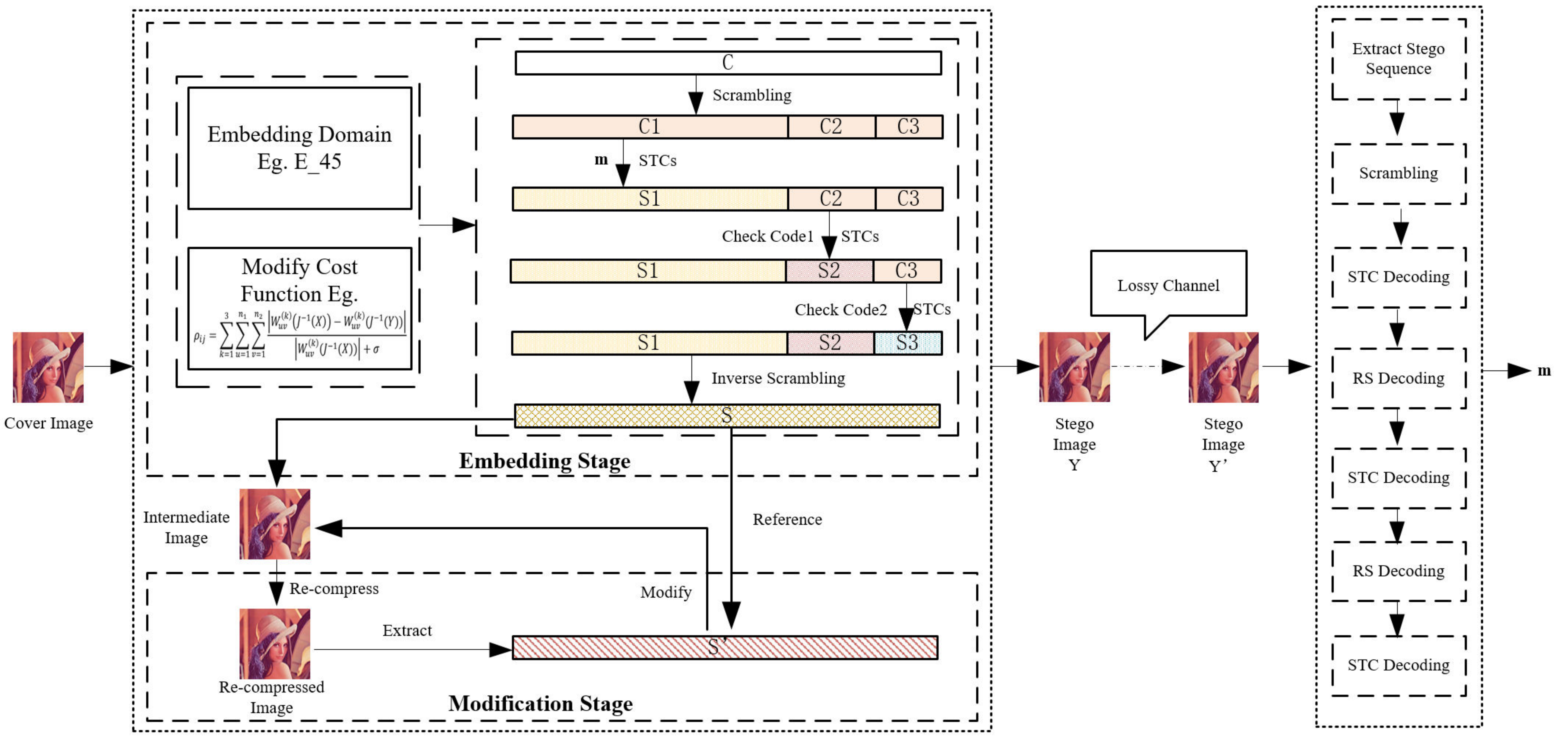}
	\caption{The framework of the proposed method}
	\label{fig4}
\end{figure*}
\par In this section, we propose ways (modification with re-compression, low frequency embedding domain, additional check codes) to improve performance of robust adaptive steganography after analyzing the embedding process of GMAS \cite{yu2020robust} and E-DMAS \cite{zhang2020enhancing}. And pseudo code of our scheme is shown in this section.

\par In the embedding process of GMAS \cite{yu2020robust} and E-DMAS \cite{zhang2020enhancing}, the de-quantized DCT coefficients are keep unchanged when the corresponding quantized DCT coefficients do not need to be modified. Some of these unmodified de-quantized DCT coefficients may be in an unstable state. For example, in Fig.3, the point $p2$ represents a unmodified de-quantized DCT coefficient, and the value of $p2$ may be affected by JPEG compression and secret messages embedding. As a possible situation, $p2$ passes through the interval to become the point $p3$, however, $p2$ represents a stego element "0" and $p3$ represents a stego element "1", which will lead to errors in STCs decoding stage. Based on this situation, we adjust the unstable DCT coefficients in the stego image with a scheme called modification with re-compression to reduce errors in the extracted stego sequences. 

\par Traditional adaptive steganography schemes have proved that embedding messages in low frequency DCT coefficients is more security than in high frequency. However, robustness of DCT coefficients in low frequency region is poor, which has been proved in GMAS \cite{yu2020robust}. Based on the modification with re-compression scheme, we need to find a new balance of security and robustness.
\par In the framework of \cite{zhang2020enhancing}, check codes of $\bm{S1}$ are embedded into the rest $l_{c}-l_{e}$ bits of the cover sequence without any error correction code. This will cause lots of errors in the STCs decoding stage if there is error in the corresponding stego sequence because of the error diffusion phenomenon of STCs. To further improve robustness of E-DMAS, we add additional check codes to the framework.
\par The flowchart of our method is shown in Fig.4. We adopt the existing steganography scheme to calculate the embedding costs of the DCT coefficients. Cover sequences are extracted from the embedding domain and divided into three segments for embedding the secret messages and check codes. After that, we can obtain intermediate images. The intermediate images are re-compressed and modified to get the final stego images. Then we transmit the stego images to the receiver with a lossy channel. The receiver extracts the secret messages from the stego images processed by the lossy channel. In the flowchart, the modification stage corresponds to modification with re-compression scheme. To facilitate the explanation, we use a pseudo code to introduce the steps of our embedding scheme in Section III E. 

\subsection{Modification with re-compression scheme} 
\par To confirm the viewpoint that part unmodified de-quantized DCT coefficients are in an unstable state, we compress 100 images selected from BOSSbase 1.01 randomly as cover images with quality factor $Q_{c}=65$. We extract cover sequences from the cover images with dither modulation when the embedding domain is E\_2345. Secret messages are embedded into the cover sequences to get stego sequences with STCs and J-UNIWARD, and we modify the cover images with the stego sequences to get stego images. Then we re-compress the stego images with channel quality factor $Q_{c}=65$ to simulate channel lossy operation. Damaged stego sequences are extracted from the re-compressed stego images and compared to the original stego sequences. Average number of errors of modified DCT coefficients and unmodified DCT coefficients in damaged stego sequences are shown in Fig.5. In addition, the average number of errors in the damaged stego sequences are shown in the Fig.6. Obviously, a large number of errors appear in the unmodified DCT coefficients, which shows that our viewpoint is correct. 
\begin{figure}	[!htb]
	\centering
	\includegraphics[width=0.45\textwidth]{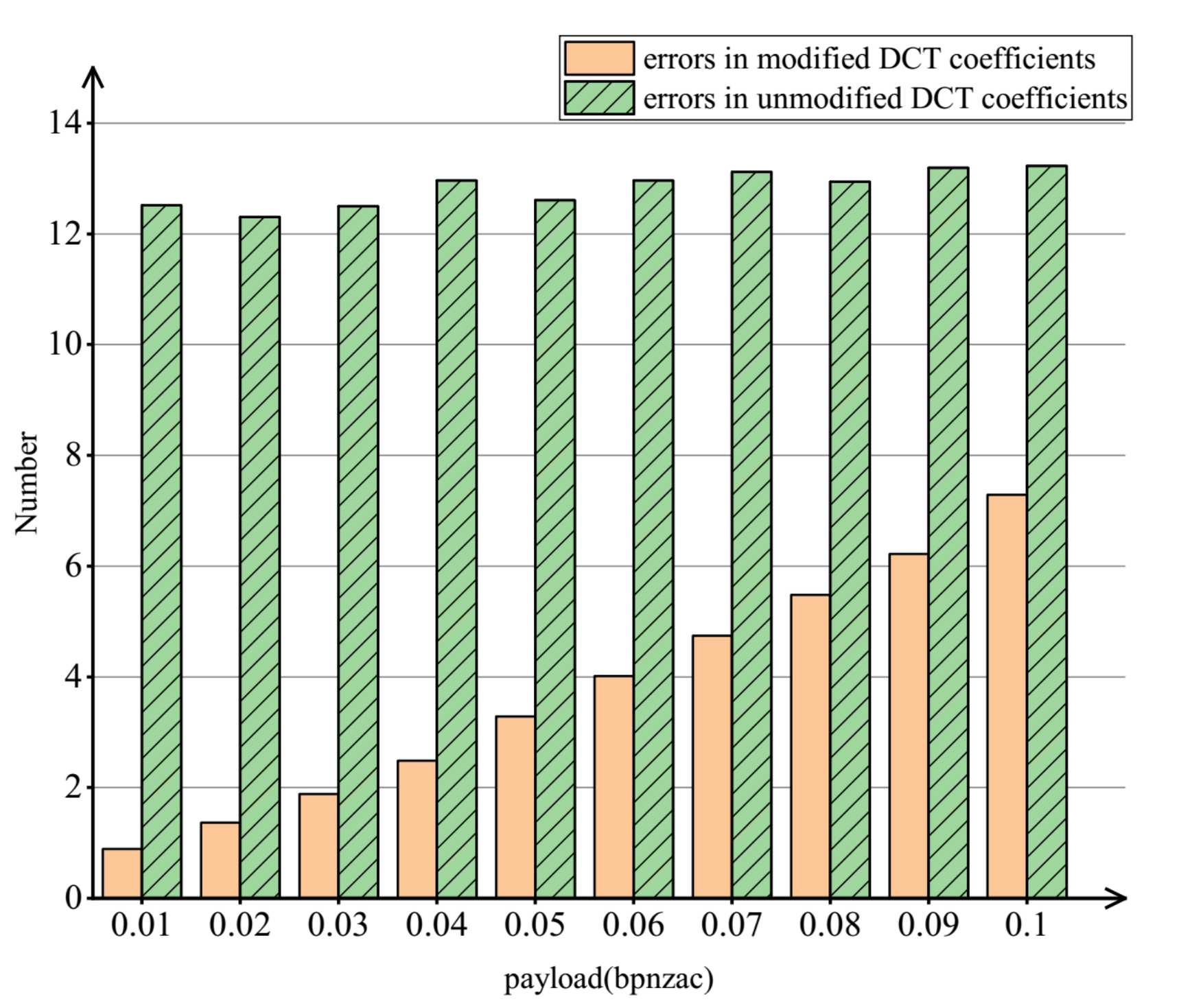}
	\caption{Average number of errors in modified DCT coefficients and unmodified DCT coefficients of damaged stego sequences on 100 images randomly selected from BOSSbase 1.01 with quality factor $Q_{c}=65$ when embedding domain is E\_2345 }
	\label{fig5}
\end{figure}
\begin{figure}	[!htb]
	\centering
	\includegraphics[width=0.45\textwidth]{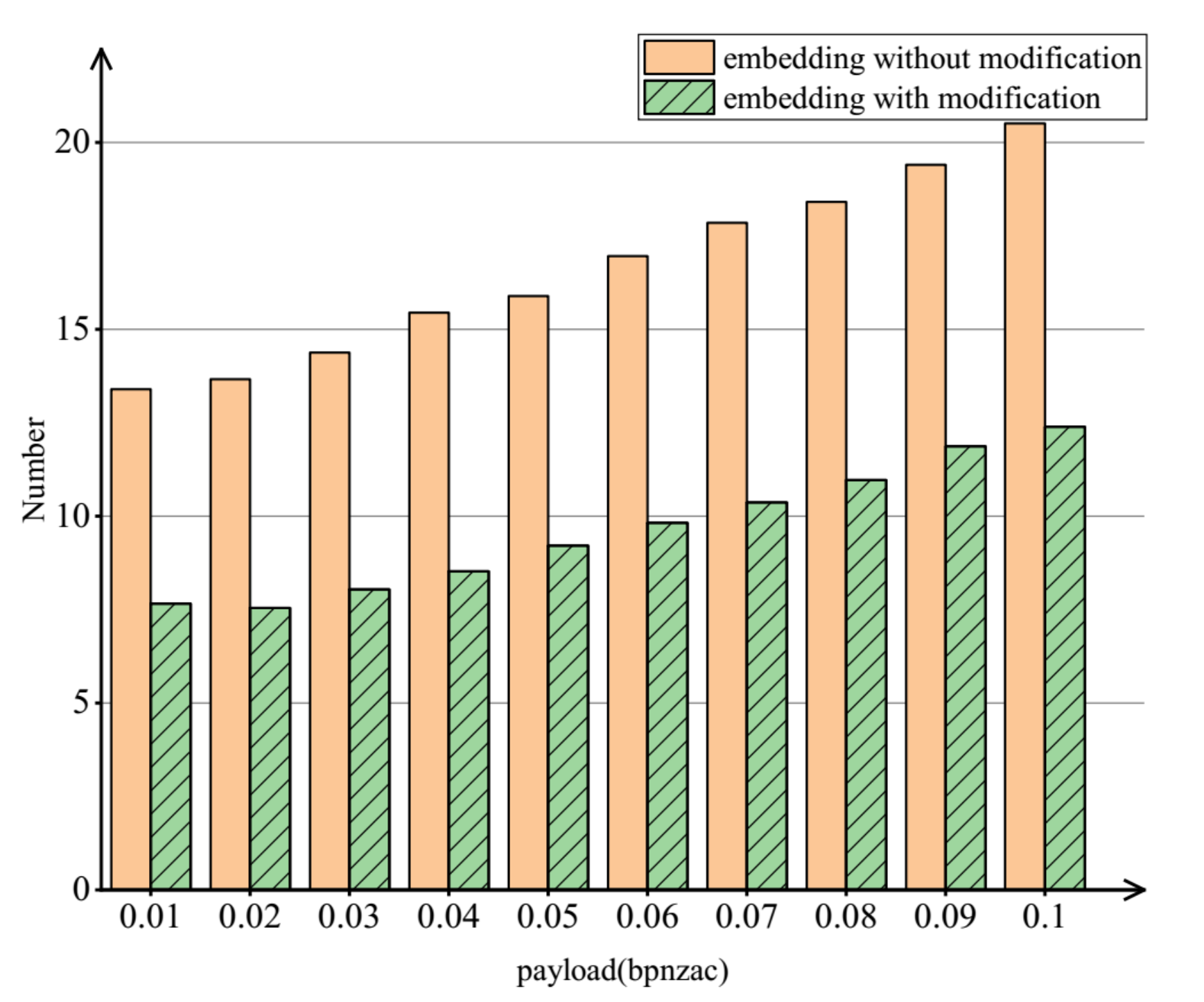}
	\caption{Average number of errors in stego sequences of embedding without modification stage or with modification stage on 100 images randomly selected from BOSSbase 1.01 with quality factor $Q_{c}=65$ when embedding domain is E\_2345  }
	\label{fig6}
\end{figure}

\par To deal with these unstable DCT coefficients, we propose a scheme called modification with re-compression. In the scheme, we re-compress stego images to cause the unstable DCT coefficients passes through intervals. After the lossy operation, we adjust these DCT coefficients to the middle coordinate of the original intervals and get the final stego images. As shown in Fig.3, the point $p2$ passes through the interval to become the point $p3$, we adjust $p3$ to the middle coordinate of the interval which $p2$ belong to. The modified value can represent the same stego element as $p2$. In order to verify the performance of this method, we add the modification with re-compression scheme to the embedding process of the above experiment. The average number of errors in the damaged stego sequences are shown in the Fig.6. It is easy to find that modification with re-compression scheme can reduce the number of errors in stego sequences. In addition, we can further improve the robustness by executing the modification with re-compression scheme multiple times which will be determined in Section IV B.

\subsection{Selection strategy of embedding domain} 
\par Although many factors can affect security of robust adaptive steganography, the impact of embedded domains is crucial. We adopt modification with re-compression scheme to improve robustness of dither modulation scheme, correspondingly, the security flaw increases because of the coefficients adjustment. Therefore, we need to balance security and robustness with moving embedding domain to low frequency regions. Besaides, contrast with GMAS \cite{yu2020robust}, the disadvantages of the embedding method and the security flaw caused by the coefficient adjustment require a lower frequency embedding domain than GMAS \cite{yu2020robust}. The modification with re-compression scheme provides the possibility of moving the embedded domain to the lower frequency regions. Constructed embedding domain is obtained with experiments in Section IV C. It is worth mentioning that current robust adaptive steganography is not suitable for embedding secret messages with high payload due to its weak security. Therefore, a smaller embedding domain is selected to reduce the number of check codes.
\subsection{Additional check codes}
\par As the description in Section II E, E-DMAS \cite{zhang2020enhancing} divide a cover sequence $\bm{C}$ into two parts, secret messages are embedded into the first part, and the second part is used to embed check codes of the sequence embedded with the secret messages. Besides, secret messages and check codes are embedded with STCs, any errors in the stego sequence will cause more errors in decoding because of the error diffusion phenomenon of STCs. It is possible that the sequence embedded with the secret messages and the corresponding check codes have errors at the same time. It is reasonable to encode the rest $l_{c}-l_{e}$ bits of the stego sequence with additional error correction code. Thus, we divide the cover sequence into three parts, first two part are similar to E-DMAS \cite{zhang2020enhancing}, and the additional check codes are embedded into the third part. The rationality of the additional check codes will be proved in Section IV D with experiments. Besides, considering that CRC codes adopted in \cite{zhang2020enhancing} have limited error correction capability. We replace CRC codes with RS codes.

\subsection{Pseudo code of our scheme}
\par In this paper, we assume that the channel quality factor $Q_{c}$ is known and adopt JPEG images with quality factor $Q_{c}$ as cover images. Our scheme is described as follows:\\
1. Investigate the quality factor $Q_{c} $ of the lossy channel and adopt JPEG images with quality factor $Q_{c}$ as cover images.\\
2. Calculate distortion $\bm{\rho}$ of a cover image with typical distortion function (e.g. J-UNIWARD) according to Eq.1.\\ 
3. Extract a cover sequence $\bm{C}$ and corresponding modification distance sequence $\bm{d}$. For all $8\times 8$ DCT blocks in the cover image, select the de-quantized DCT coefficients in E\_45 to calculate the cover sequence $\bm{C}$ and modification distance sequence $\bm{d}$ with dither modulation algorithm.\\
4. Calculate the modification costs $\bm{\xi}$. $\rho_{ij}$ represents the cover image distortion of  $ij$th quantized DCT coefficient. According to Eq.6, $\zeta_{ij}$ represents the cover image distortion of  $ij$th de-quantized DCT coefficient. We can calculate modification costs $\bm{\xi}$ with Eq.6.\\
5. Scramble the cover sequence $\bm{C}$ to get a sequence $\bm{C'}$. We divide the sequence $\bm{C'}$ into 3 segments ($\bm{C1}, \bm{C2}, \bm{C3}$) to embed a secret message, check codes and additional check codes respectively, in order to balance the payload of each segment, we set $l_{C1}:l_{C2}:l_{C3} \approx 15:3:1$ according to the ratio of the average length of the secret messages, the length of the check codes, and the length of the additional check codes. ($l_{C1}$ represents the length of $\bm{C1}$).\\
6. Embed a secret message $\bm{m}$ into the sequence $\bm{C1}$ with STCs to get a stego sequence $\bm{S1}$.\\
7. Encode the stego sequence $\bm{S1}$ with RS codes and embed the check codes of $\bm{S1}$ into sequence $\bm{C2}$ with STCs to get a stego sequence $\bm{S2}$.\\
8. Encode the stego sequence $\bm{S2}$ with RS codes and embed the check codes of $\bm{S2}$ (called additional check codes later) into sequence $\bm{C3}$ with STCs to get a stego sequence $\bm{S3}$. Finally, we get a sequence $\bm{S}$ after inverse scramble the stego sequence ($\bm{S1}, \bm{S2}, \bm{S3}$).\\
9. Modify de-quantized DCT coefficients in embedding domain of the cover image according to the sequence $\bm{S}$ with dither modulation algorithm, so that the sequence extracted from the modified de-quantized DCT coefficients is consistent with the sequence $\bm{S}$. Intermediate image $\bm{I}$ can be obtained with the modified DCT coefficients.\\
10. Re-compress the intermediate image $\bm{I}$ with quality factor $Q_{c} $ to cause the unstable DCT coefficients change,
and extract a stego sequence $\bm{S'}$ from the re-compressed intermediate image $\bm{I'}$ with the dither modulation algorithm.\\
11.	Compare $\bm{S'}$ with $\bm{S}$, if there is difference in $\bm{S'}$, we modify the corresponding de-quantized DCT coefficient in $\bm{I'}$ with dither modulation algorithm.\\
12.	Repeat step 10 and step 11 once to further reduce errors in the extracted sequence and get a final stego image $\bm{Y}$.
\par The receiver calculates the de-quantized DCT coefficients of the received stego image, then the stego sequence $\bm{S}$ is extracted with dither modulation algorithm and scrambled. Additional check codes are extracted from $\bm{S3}$ with STCs decoding to correct the $\bm{S2}$. Similarly, check codes of $\bm{S1}$ are extracted from corrected $\bm{S2}$ with STCs decoding to correct the $\bm{S1}$. After that, the secret message is extracted from corrected $\bm{S1}$ sequence.

\section{Experimental results and discussion}
\par In this part, we obtain the appropriate parameters of our proposed scheme with experiments at first. Then our scheme is compared with GMAS\cite{yu2020robust} and E-DMAS \cite{zhang2020enhancing} in security and robustness. 
\subsection{Setups}
\label{sec::Conclusion}
\par The dataset of all experiments conducted in this paper is BOSSbase 1.01\cite{bas2011break} which contain 10000 512*512 grayscale images. We assume that the channel quality factors are $Q_{c}=65,75$. We compress these images with quality factor $Q_{c}$. We set the parameter $h=10$ of STCs just as GMAS\cite{yu2020robust} and E-DMAS\cite{zhang2020enhancing}. The range of payloads adopted in this paper is 0.01 to 0.1bpnzac (bits per non-zero AC coefficients). It is worth mentioning that we treat every 8-bit stego sequence as an integer and encode the integer sequence with RS (255,251). Besides, errors in a stego sequence exceed the error correction ability of RS codes, it will not be corrected. And we believe that at most one bit in each integer is wrong because of the strong robustness of the cover sequence.  
\par In the following experiment, 1000 images are selected randomly from BOSSbase 1.01 \cite{bas2011break} as original images. We compress these original images with quality factor $Q_{c}$ to get cover images. We embed secret messages into cover images to get stego images, then the stego images are re-compressed with the quality factor $Q_{c}$ to simulate channel lossy operation. We extract the secret messages from the re-compressed stego images and calculate average extraction error rates of the secret messages. The average extraction error rates of secret messages are used to evaluate the robustness of schemes.
\par For security, we conduct experiments on whole 10000 images of BOSSbase 1.01 \cite{bas2011break}. We utilize CCPEV (Typical Cartesian Calibrated PEV) \cite{kodovsky2009calibration} and DCTR (Discrete Cosine Transform Residual) \cite{6935011} algorithms to extract features of cover and corresponding stego images. We divide the extracted features into a training set and a testing set. The ensemble classifier\cite{6081929} is trained and tested respectively when the ratio of the training set to the testing set is 1:1 to 10:1, the average of 10 test results is used as the detection error rate.
\subsection{Performance of modification with re-compression scheme}
\begin{figure}	[!htb]
	\centering
	\includegraphics[width=0.45\textwidth]{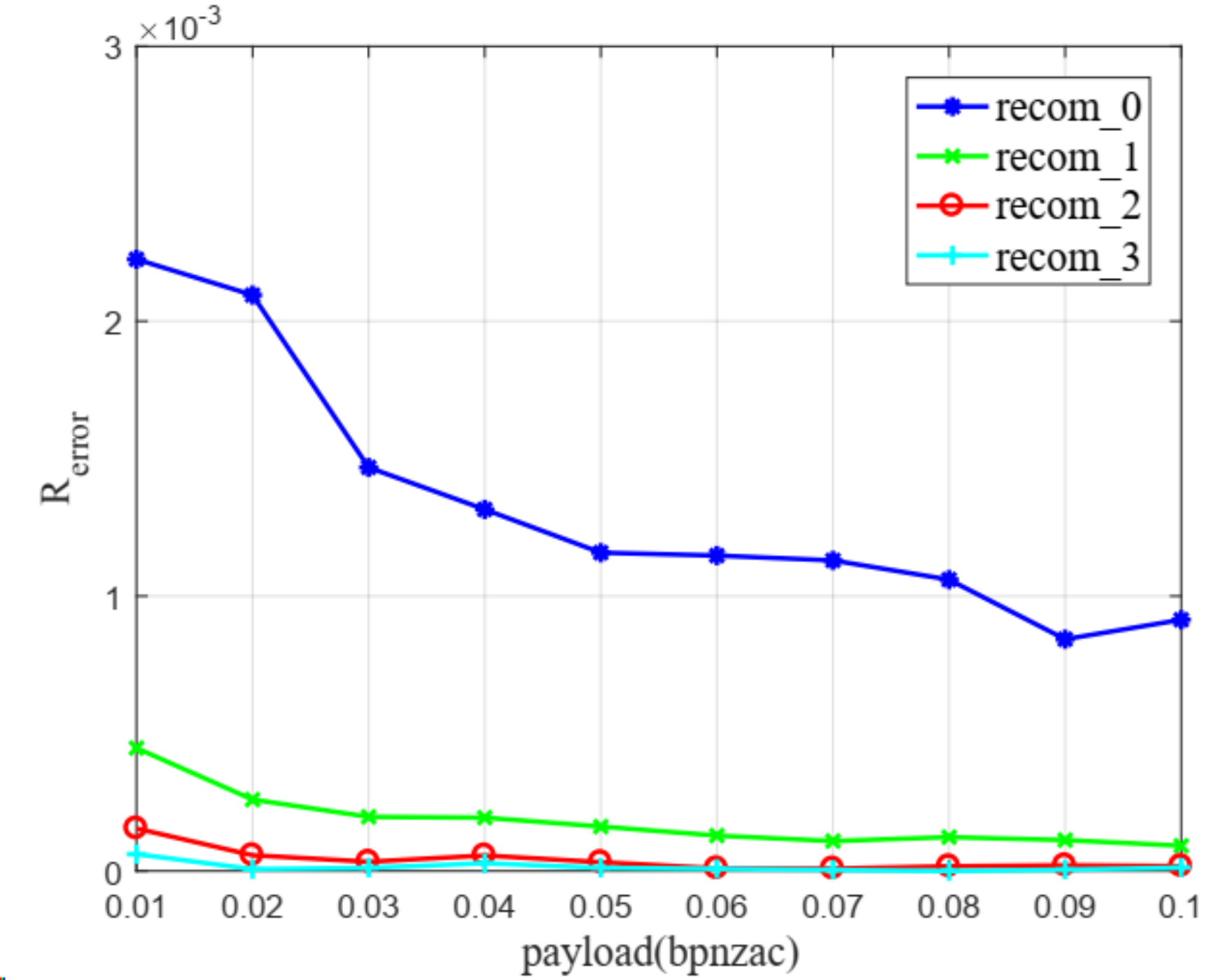}
	\caption{Average extraction error rates of secret messages after performing 0,1,2,3 times modification with re-compression on 1000 images randomly selected from BOSSbase 1.01 with quality factor $Q_{c}=65$ when embedding domain is E\_45 }
	\label{fig7}
\end{figure}
\par In Section III B, we proved that the modification with re-compression scheme is effective in improving the robustness of  steganography. Obviously, the  modification stage as shown in Fig.4 can be executed multiple times to achieve better results, we conducted a comparative experiment with $Q_{c}=65$ to determine the execution number of modification stage. The intermediate images can be obtained by embedding secret messages and check codes. We execute the modification stage 0, 1, 2, 3 times and named "recom\_0", "recom\_1", "recom\_2", "recom\_3" correspondingly to get the final stego images. Then, the stego images are re-compressed with quality factor $Q_{c}$ to simulate channel lossy operation. Finally, secret messages are extracted from the re-compressed stego images to calculate the average extraction error rates. The average extraction error rates of secret messages under different number of executing modification stage are shown in Fig.7. With the number of executing  modification stage increase, the average extraction error rates of secret messages gradually decreases. When we execute the modification stage more than twice, improvement in terms of robustness is not obvious, and it will greatly increase the running time and risk of being detected. Therefore, we decided to execute the modification stage twice for each intermediate image.
\subsection{Performance of constructed embedding domain}
\begin{figure}[!htb]	
	\centering
	\includegraphics[width=0.45\textwidth]{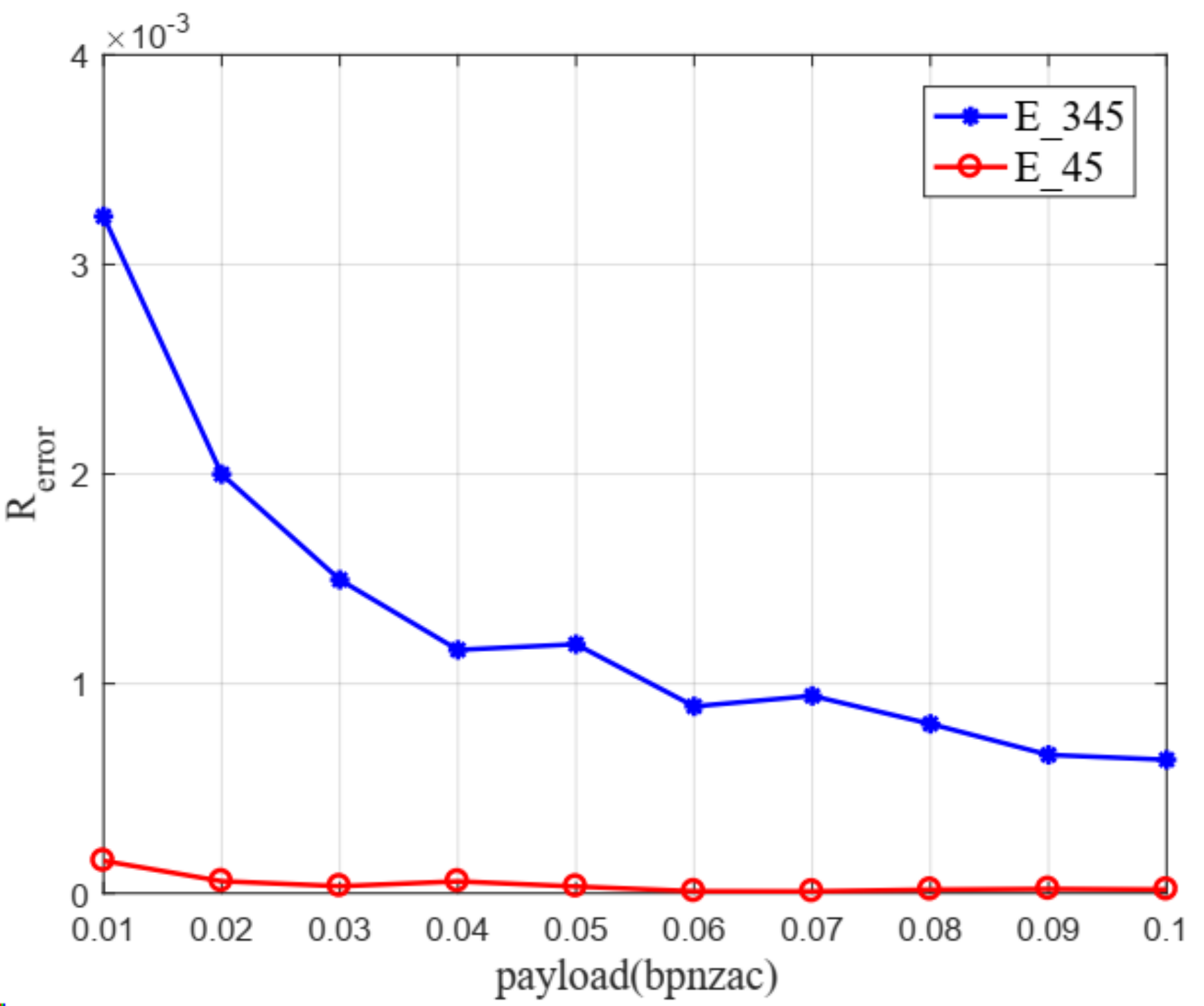}
	\caption{Average extraction error rates of secret messages with different embedding domain after performing twice modification with re-compression on 1000 images randomly selected from BOSSbase 1.01 with quality factor $Q_{c}=65$ }
	\label{fig8}
\end{figure}
\par We try to improve the security of robust adaptive steganography with embedding secret messages in low frequency regions. However, robustness of DCT coefficients in low frequency regions is poor. Fortunately, modification with re-compression scheme can reduce the error rates of extracted stego sequences, which allows us to embed secret messages into the low frequency regions. Just as describe in Section III C, we adopt lower frequency embedding domain than GMAS \cite{yu2020robust} to guarantee security of robust adaptive steganography. We adopt the proposed scheme to generate stego images with different embedding domains when quality factor $Q_{c}=65$. E\_45 is selected as embedding domain at first and expanded to the lower frequency regions gradually. We embed secret messages and check codes into cover images based on these different embedding domains to get stego images. Then the stego images are re-compressed with quality factor $Q_{c}$ to simulate channel lossy operation. As shown in Fig.8, the average extraction error rates of secret messages is calculated with extracted secret messages. The average extraction error rates of the secret messages are much higher when the embedding domain is E\_345. Yu et al. have proved in \cite{yu2020robust} that the lower frequency the embedding domain, the weaker the robustness of cover sequences. Therefore, we do not further expand the embedding domain to lower frequency regions. E\_45 is selected as embedding domain in this paper.
\subsection{Performance of additional check codes}
\begin{figure}[!ht]	
	\centering
	\includegraphics[width=0.45\textwidth]{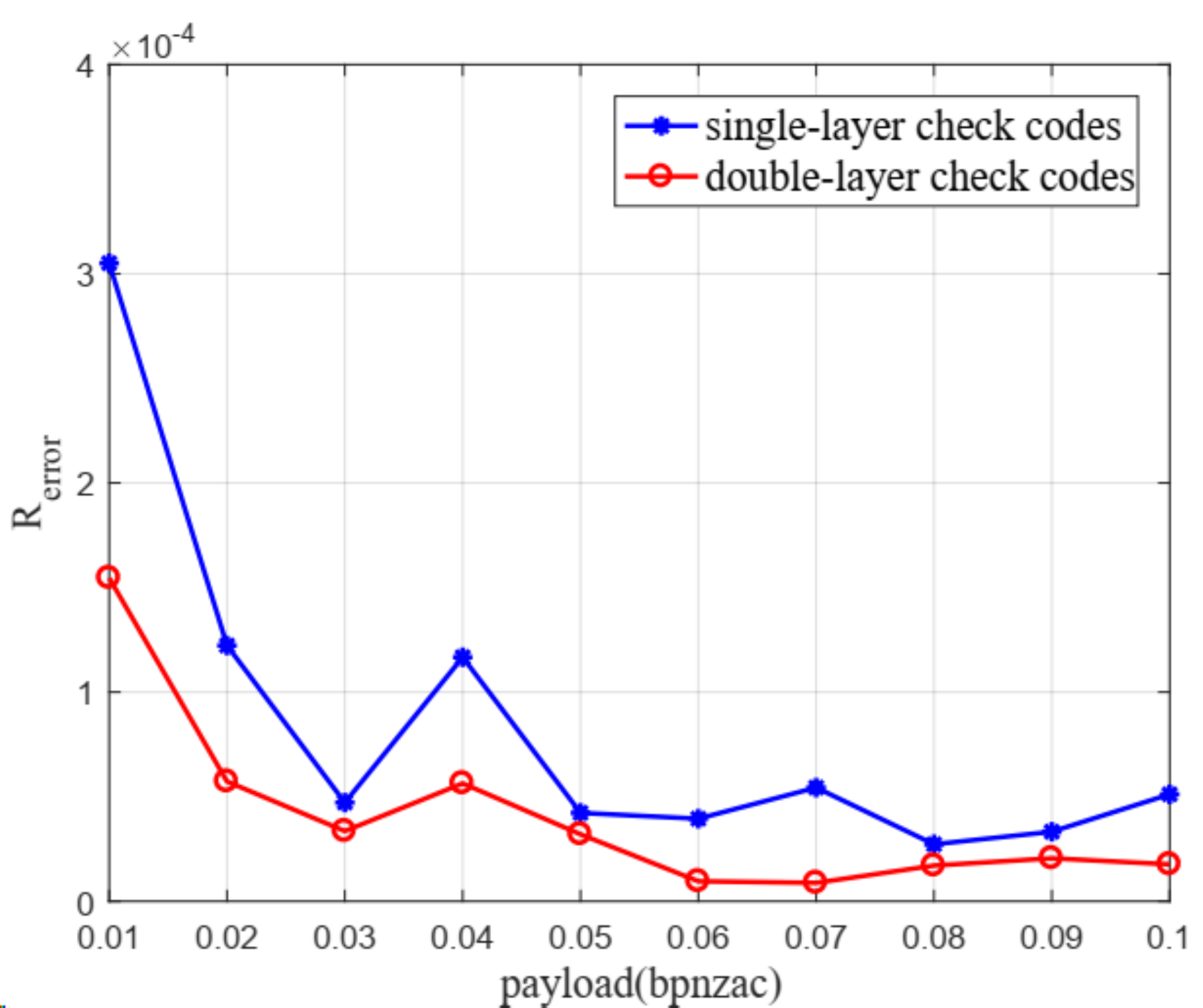}
	\caption{ Average extraction error rates of secret messages with single-layer or double-layer check codes after performing twice modification with re-compression on 1000 images randomly selected from BOSSbase 1.01 with quality factor $Q_{c}=65$ when embedding domain is E\_45 }
	\label{fig9}
\end{figure}
\par As described in Section III D, since the check codes of E-DMAS \cite{zhang2020enhancing} are embedded into the rest $l_{c}-l_{e}$ bits cover sequence, which is not robust. To solve such problem, based on the original framework, we propose additional check codes which corresponds to step 8 in Section III E. To verify the performance of additional check codes, we conduct an experiment with $Q_{c}=65$. We represent the scheme without additional check codes as "single-layer check codes", otherwise "double-layer check codes". We embed secret messages and check codes with single-layer check codes or double-layer check codes respectively to generate intermediate images. Then, we execute the modification stage twice for the intermediate images to generate stego images. The stego images are re-compressed with quality factor $Q_{c}$ to simulate channel lossy operation. As shown in Fig.9, the average extraction error rates of the scheme with or without additional check codes is calculated with extracted secret messages. The results demonstrate that additional check codes are useful for reducing average extraction error rates of secret messages.
\subsection{Compare our scheme with E-DMAS and GMAS in robustness}
\begin{figure*}[!ht]
	\centering
	\begin{minipage}[t]{0.45\textwidth}
		\centering
		\subfigure[]{
			\includegraphics[width=\textwidth]{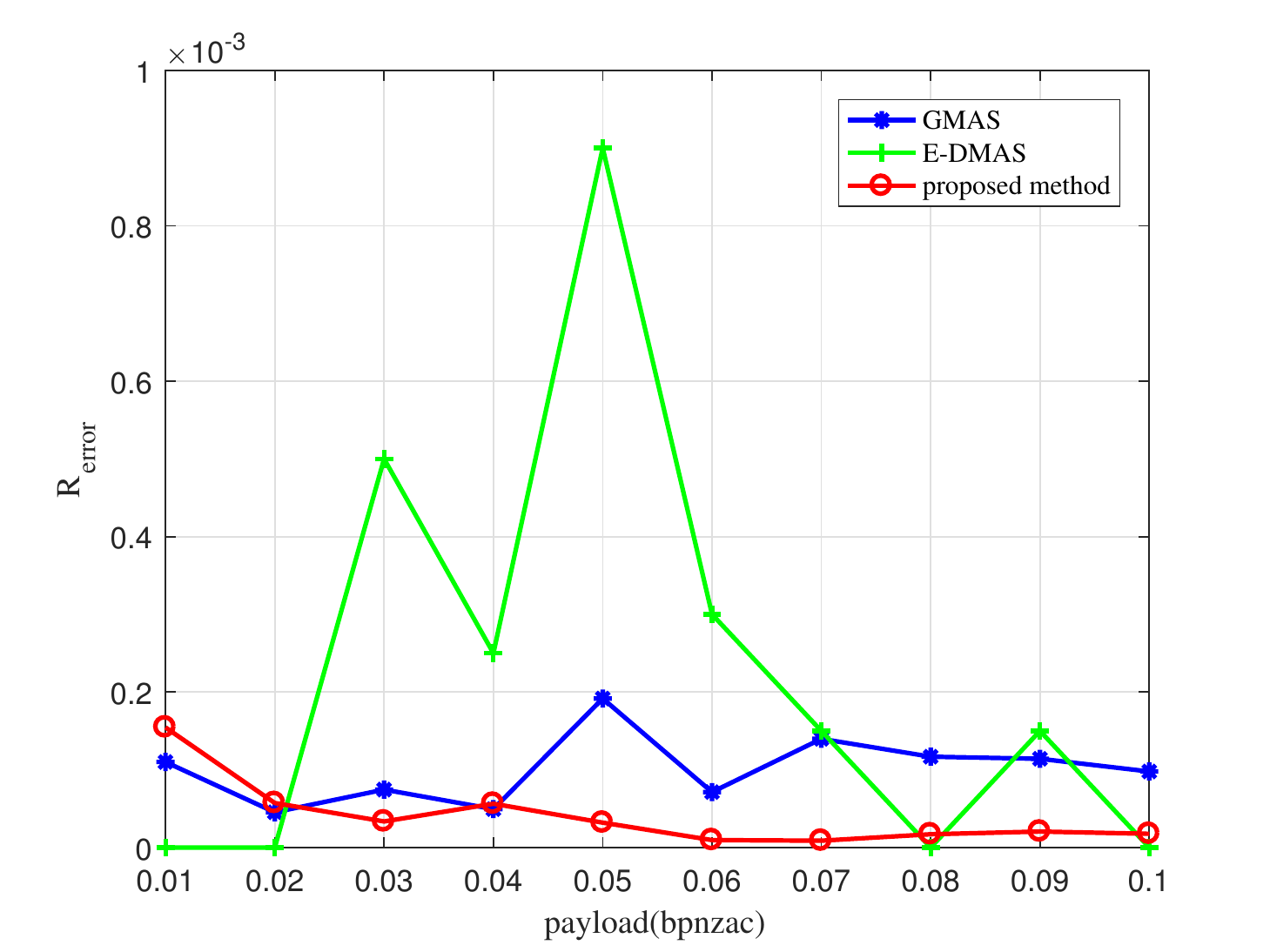}
		}
	\end{minipage}
	\begin{minipage}[t]{0.45\textwidth}
		\centering
		\subfigure[]{
			\includegraphics[width=\textwidth]{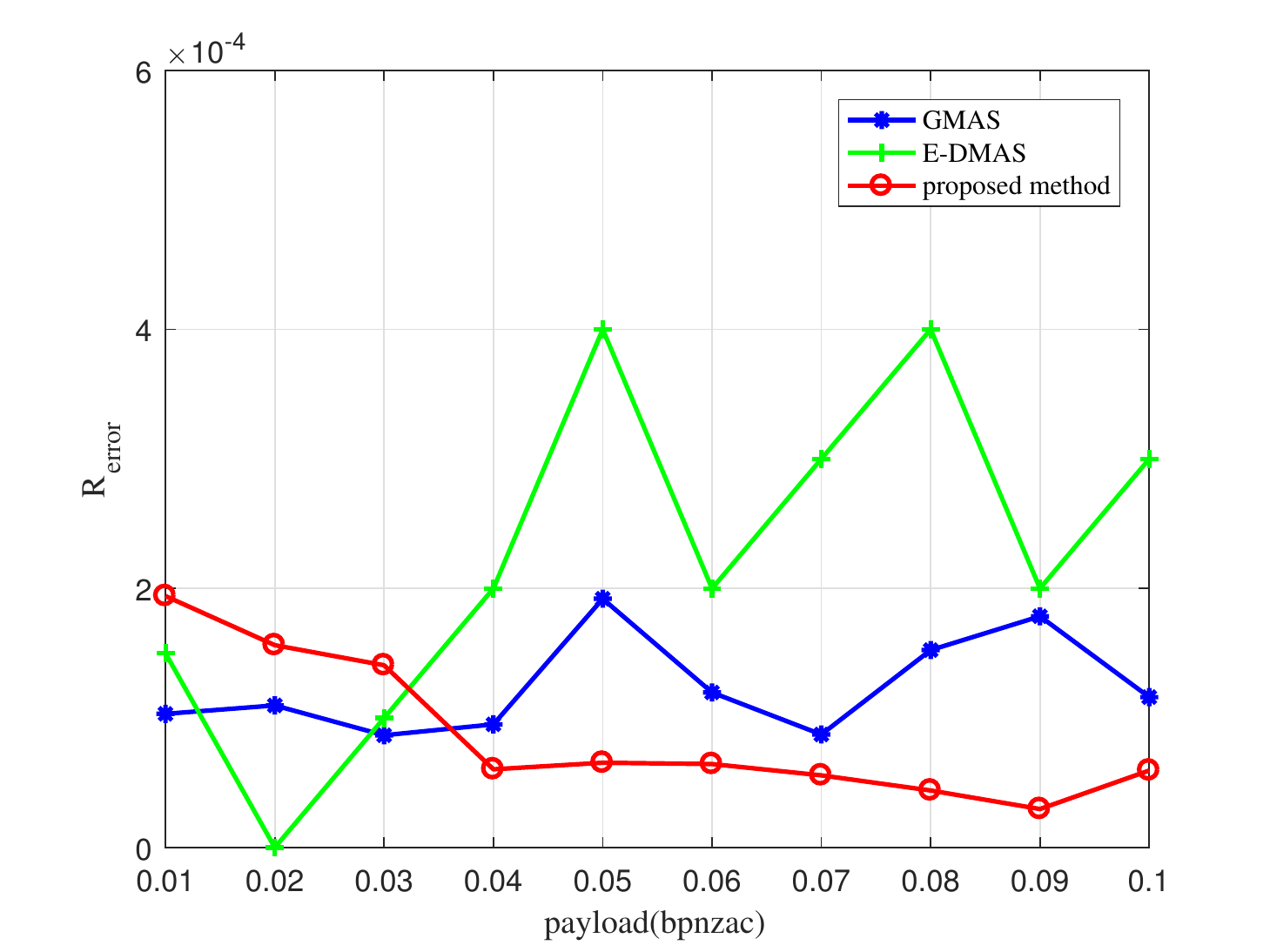}
		}
	\end{minipage}
	\caption{Average extraction error rates of GMAS\cite{yu2020robust}, E-DMAS\cite{zhang2020enhancing} and proposed method after performing twice modification with re-compression when embedding domain is E\_45 with channel quality factor $Q_{c}=65$(left) and $Q_{c}=75$(right) on 1000 images randomly selected from BOSSbase 1.01 }	
	\label{fig10}
\end{figure*}
\begin{figure*}[!ht]
	\centering
	\begin{minipage}[t]{0.45\textwidth}
		\centering
		\subfigure[]{
			\includegraphics[width=\textwidth]{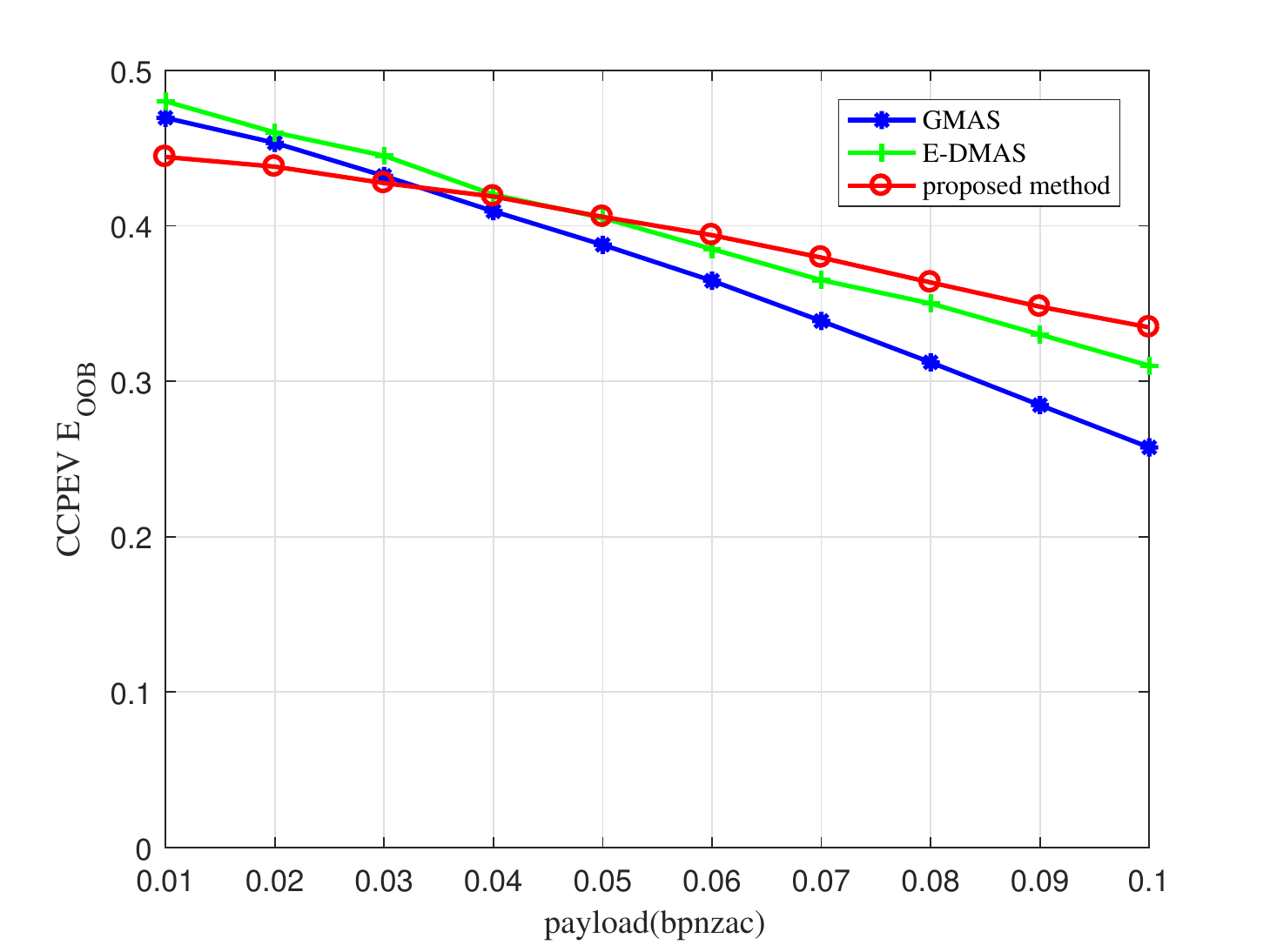}
		}
	\end{minipage}
	\begin{minipage}[t]{0.45\textwidth}
		\centering
		\subfigure[]{
			\includegraphics[width=\textwidth]{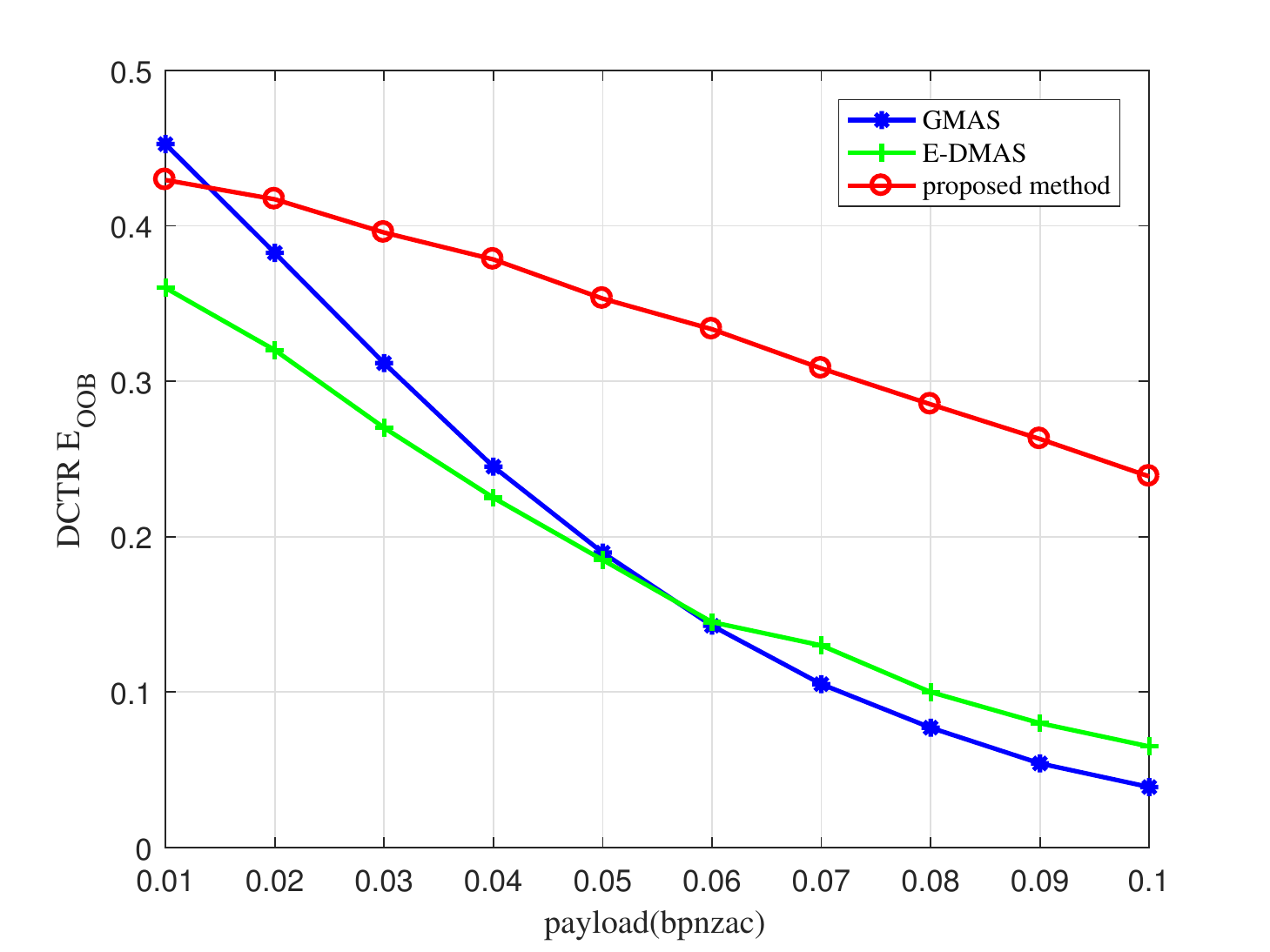}
		}
	\end{minipage}
	\caption{Average detection error rates of GMAS\cite{yu2020robust}, E-DMAS\cite{zhang2020enhancing} and proposed method after performing twice modification with re-compression when embedding domain is E\_45 against CCPEV(left) and DCTR(right) with channel quality factor $Q_{c}=65$ on 10000 images selected from BOSSbase 1.01}
\end{figure*}
\par In this part, we will compare our scheme with GMAS \cite{yu2020robust} and E-DMAS \cite{zhang2020enhancing} in robustness. We conduct experiments with channel quality factor $Q_{c}=65,75 $ correspondingly. We extract the secret messages from the damaged stego images, the average extraction error rates of secret messages are shown in Fig.10. Obviously, our scheme can achieve the same and even stronger robustness than E-DMAS \cite{zhang2020enhancing} and GMAS\cite{yu2020robust} when payloads larger than 0.03bpnzac. Although it performs poorly at low payloads, but not far from the comparative schemes.
\par Because of the strong robustness of the middle frequency regions, E-DMAS adopts a small number of check codes for error correction, once the extracted stego sequences have uncorrectable errors, a lot of errors will appear in the decoded secret messages because of the error diffusion phenomenon of STCs. Thus, the robustness of E-DMAS \cite{zhang2020enhancing} fluctuates greatly. As shown in Fig.5, with the amount of embedded secret messages increases, errors in extracted stego sequences increase gradually. GMAS \cite{yu2020robust} adopts ternary STCs and the way which encodes secret messages with error correction code. Thus, with payload increase, the possibility that the number of errors in the decoded secret messages exceeds the error correction capability increase. To a large extent, our scheme can solve the increasing number of errors in the extracted stego sequences with payload increase, which cause the number of errors of decoded secret messages within a certain range. When it is low payload, the amount of secret messages is small, which lead to the higher average extraction error rates. As the payload increases, the average extraction error rates gradually decrease.
\subsection{Compare our scheme with E-DMAS and GMAS in security}
\par As for security, we conduct experiments on the entire Bossbase 1.01 with quality factor $Q_{c}=65 $. Steganalysis is essentially a binary classification to distinguish cover images and stego images. CCPEV \cite{kodovsky2009calibration} and DCTR \cite{6935011} steganalysis algorithms are utilized to extract features of cover and stego images. The ensemble classifier is trained with the extracted features. The classification error rates of steganalysis with different payloads are shown in the Fig.11. Obviously, the high classification error rate indicates that the steganalysis algorithm cannot distinguish the cover images and the corresponding stego images, which means the security of the steganography algorithm is high. Compared with E-DMAS \cite{zhang2020enhancing} and GMAS \cite{yu2020robust}, our scheme greatly improve security in terms of DCTR features \cite{6935011}. However, improvement in resist CCPEV \cite{kodovsky2009calibration} detection ability is not obvious. The experiment results show that when the payloads are larger than 0.03bpnzac, the resist detection ability is improved. At low payloads, the security is not far from the comparative schemes. 
\par GMAS \cite{yu2020robust} adopt ternary STCs which have better security performance than binary STCs and the way which encode secret messages with error correction code, the amount of check codes is proportional to the amount of secret messages. Because of the strong robustness of the middle frequency regions, E-DMAS \cite{zhang2020enhancing} adopts a small number of check codes for error correction. When it is low payload, effective embedding methods and fewer check codes lead to better performance of the comparison schemes. As the payload increases, the advantages of framework and low frequency embedding domains gradually appear.
\section{Conclusion}
\par Nowadays, social networks are more and more widely used in our lives, which provides the possibility for steganography. Because of the lossy process of social networks, such as JPEG re-compression, the robustness of adaptive steganography needs to be improved. 
\par In this paper, we propose a scheme called modification with re-compression to improve robustness of E-DMAS\cite{zhang2020enhancing}. And we move embedding domain to lower frequency region than GMAS \cite{yu2020robust} to balance robustness and security. In addition, we add additional check codes to improve robustness. The experiment results demonstrate our scheme can achieve higher security and robustness than comparative schemes when payloads are larger than 0.03bpnzac. The security and robustness are not far from the comparative papers when the payloads are less than 0.03bpnzac.  
\par In the future, we will look for solutions that can improve security without prior knowledge of channel quality factor. Another future work is to increase embedding capacity.
\section*{Acknowledgments}
\label{sec::Acknowledgments}
This research work is partly supported by National Natural Science Foundation of China (61872003, U1636206). State Key Laboratory of Computer Architecture (ICT,CAS) under Grant No. CARCHB202018.
\ifCLASSOPTIONcaptionsoff
  \newpage
\fi



%

\bibliographystyle{IEEEtran}
\bibliography{ref.bib}

%








\end{document}